\documentclass[11pt]{article}

\usepackage{amsmath,amssymb,color,graphicx,cite}
\usepackage{titlesec}
\usepackage[titletoc,title]{appendix}
\usepackage{amsthm}
\usepackage{subcaption} 
\usepackage{placeins}

\newcommand{\hoch}[1]{$\, ^{#1}$}

\usepackage[normalem]{ulem}

\newcommand{\be}{\begin{equation}}
\newcommand{\ee}{\end{equation}}
\newcommand{\bea}{\setlength\arraycolsep{2pt} \begin{eqnarray}}
\newcommand{\eea}{\end{eqnarray}}

\def\ft#1#2{{\textstyle{\frac{\scriptstyle #1}{\scriptstyle #2} } }}
\def\fft#1#2{{\frac{#1}{#2}}}

\def\0{{\sst{(0)}}}
\def\1{{\sst{(1)}}}
\def\2{{\sst{(2)}}}
\def\3{{\sst{(3)}}}
\def\4{{\sst{(4)}}}
\def\5{{\sst{(5)}}}
\def\6{{\sst{(6)}}}
\def\7{{\sst{(7)}}}
\def\8{{\sst{(8)}}}
\def\sst#1{{\scriptscriptstyle #1}}

\begin{document}

\begin{center}
{\Large {\bf Classification of Oppenheimer-Snyder Collapse: Singular, Bouncing, and Soft-Landing Scenarios}}

\vspace{20pt}
Zhi-Chao Li\hoch{1}, H.~Khodabakhshi\hoch{1} and H.~L\"u\hoch{1,2}

\vspace{10pt}

{\it \hoch{1}Center for Joint Quantum Studies, Department of Physics,\\
School of Science, Tianjin University, Tianjin 300350, China }

\medskip

{\it \hoch{2}The International Joint Institute of Tianjin University, Fuzhou,\\ Tianjin University, Tianjin 300350, China}

\vspace{40pt}

\end{center}

\begin{abstract}
	We study Oppenheimer-Snyder (OS) gravitational collapse matched to a general static, spherically symmetric exterior spacetime. Unlike the Schwarzschild case, two new features can arise in black holes with two horizons: an \emph{apparent-horizon minimum}, a temporary minimum in the apparent-horizon radius during collapse, and a \emph{bounce}, where the star surface stops collapsing at a nonzero radius and reverses into expansion. We identify the conditions that lead to these two features. For two-horizon exteriors, trapped-region consistency requires that the apparent-horizon turning point occurs no earlier than the surface crossing of the inner horizon. As a concrete example, the OS collapse of the Reissner-Nordstr\"om (RN) spacetime shows both effects. In contrast, regular black holes with de Sitter cores show neither: their collapse is smooth and monotonic, and the surface approaches the center only as the proper time goes to infinity. These results naturally classify the OS collapses into three categories: \emph{singular}, which ends at the center in finite time; \emph{bouncing}, which reverses at a finite radius; and \emph{soft-landing}, which reaches the center only asymptotically. We argue that these features are consistent with Penrose's strong cosmic censorship conjecture.

\end{abstract}

\vfill{\footnotesize lizc@tju.edu.cn \ \ \  h\_khodabakhshi@tju.edu.cn
\ \ \ mrhonglu@gmail.com}

\thispagestyle{empty}
\pagebreak

\setcounter{tocdepth}{2}
\tableofcontents
\pagebreak

\section{Introduction}
\label{sec:intro}

Gravitational collapse is one of the most important predictions of General Relativity and provides the theoretical foundation for black hole formation. The classic Oppenheimer-Snyder (OS) model \cite{OppenheimerSnyder:1939} describes how a homogeneous, pressureless dust star collapses into a Schwarzschild black hole. The interior Friedmann-Lema\^{\i}tre-Robertson-Walker (FLRW) spacetime is smoothly matched to the exterior Schwarzschild metric; in modern language, this matching satisfies the Israel junction conditions \cite{IsraelJunction:1966}. In this idealized picture, the star surface reaches the central spacelike singularity in finite proper time, while trapped surfaces form inside the star. The boundary of the trapped region within the interior is the inner apparent horizon. For brevity, we refer to this simply as the apparent horizon. It shrinks along with the star surface to zero size.

There are additional black holes beyond the Schwarzschild black hole. A notable example is the Reissner-Nordstr\"om (RN) black hole. It is thus of interest to study generalized OS collapse such that the exterior of the collapsing star surface is described by a general spherically-symmetric black hole metric, namely
\be
ds^2=-h(r) dt^2 + \fft{dr^2}{f(r)} + r^2 d\Omega_2^2\,.\label{genstatic}
\ee
However, it turns out that the consistency conditions on the matching between the interior FLRW spacetime and the exterior static metric \eqref{genstatic} require that $h=f$ \cite{OurProof}. In other words, the generalized OS (GOS) collapse beyond the Schwarzschild black hole applies only to the special static metric with $g_{tt} g_{rr}=-1$. Nevertheless, there still exist large classes of static black holes that are of the special static type, including the RN black holes and most of the regular black holes, e.g.~\cite{bardeen,Hayward:2006}. In this framework, the interior matter near the star surface is modeled as neutral pressureless dust in geodesic freefall. The electric charge is assumed to be carried by the star surface $\Sigma$, contributing to the electromagnetic surface current. This should be distinguished from a gravitational thin shell, which is characterized by a nonzero surface stress-energy tensor. In the present construction, the induced metric and the extrinsic curvature are continuous across $\Sigma$. The FLRW bulk interior is neutral and contains only pressureless dust in geodesic free fall. Therefore, no radial electromagnetic field is present in the FLRW bulk. Its electromagnetic field permeates the exterior spacetime, producing a radiation-like term ($\rho_{\rm EM} \propto R^{-4}$) in the effective energy density evaluated at the surface \cite{OurPRD}. As we will show later, this approach is justified by two important aspects. The first is that it satisfies both Einstein's and Maxwell's equations at the surface, ensuring a valid matching via the Israel junction conditions, and the second is that it turns out that the assumed neutral star surface will bounce back after passing the inner horizon. Since the surface reaches a turning point at $R_\ast>0$, the OS trajectory never probes the central singular region $R=0$. Consequently, within the present matching framework, the bounce is determined solely by the exterior metric function $f(R)$ evaluated at the surface. In our recent work \cite{OurPRD}, we generalized OS collapse to an arbitrary special static and spherically-symmetric exterior. We derived general formulae for the surface evolution, as well as for both the apparent and event horizons. We also identified an upper bound on the event-horizon formation time and conjectured that Schwarzschild collapse saturates this bound for fixed mass. We later proved this conjecture in \cite{OurProof}, assuming that the matter satisfies the weak energy condition.

Both RN and regular black holes are characterized by having two horizons, the outer event horizon and inner Cauchy horizon. Extending the OS framework to such exteriors can reveal new features. We illustrate this using the RN black hole as a concrete example. In our previous work \cite{OurPRD}, we noticed that the repulsive effect inside the inner horizon can stop the collapse at a nonzero radius, though we did not identify it as a bouncing point. We also used the apparent horizon structure to constrain the charge. However, we did not formally define these features or derive general criteria for them.

A closely related construction was developed in the quantum Oppenheimer-Snyder (qOS) and quantum Swiss Cheese models of Ref.~\cite{Lewandowski:2023}. In that work, the classical dust FLRW interior is replaced by the effective Ashtekar-Pawlowski-Singh dynamics of loop quantum cosmology (LQC). The junction conditions then lead to a quantum-deformed Schwarzschild exterior of the special-static form, with 
$
	f_{\text{qOS}}(r) = 1 - \frac{2m}{r} + \frac{\alpha  m^2}{r^4},
$
where $m$ is the ADM mass and $\alpha$ is a parameter proportional to the square of the Planck length. The quantum correction produces a minimum radius for the dust surface, 
$
	r_b = \left( \frac{\alpha m}{2} \right)^{1/3},
$
and implies a lower mass bound for black hole formation. For masses above this bound, the exterior exhibits a two-horizon structure analogous to the RN spacetime, while the dust surface reaches $r_b$ and bounces.  The present work is complementary to that construction. Rather than assuming a specific LQC effective Friedmann equation, we start from a general special-static exterior metric $f(r)$ and derive purely geometric criteria for a star-surface bounce and an apparent-horizon minimum. The qOS metric of Ref.~\cite{Lewandowski:2023} therefore serves as a concrete, physically motivated example to which our general criteria can be applied. In this paper, we focus on identifying new physically meaningful signatures in OS collapse dynamics by considering the more general non-Schwarzschild exterior metric function $f(r)$. We reveal two independent features that can arise: (i) a \emph{bounce}, where the star's surface reaches a nonzero finite radius and reverses from collapse to expansion; and (ii) an \emph{apparent-horizon minimum}, where the apparent-horizon radius reaches a temporary minimum during the collapse. For a physically consistent formation of a two-horizon black hole, the apparent-horizon turning point must occur no earlier than the surface crossing over the inner horizon, so that the relevant apparent-horizon branch connects monotonically between the outer horizon $R_+$ and the inner horizon $R_-$ \cite{OurPRD}. This condition provides a strong constraint on the black hole parameters. In the RN case, it implies a lower bound on the charge, namely $|q|\ge \sqrt{3}\,m/2$. Importantly, this also ensures that the bounce radius $R_\ast$ is not negligible, making the bounce a physically meaningful feature. In addition, we introduce a third characteristic scale, namely the \emph{inflection radius} \(R_{\rm infl}\), where the surface acceleration vanishes (\(\ddot R=0\)) while the collapse is still ongoing. This radius marks the transition from the accelerating phase to the decelerating phase of the collapse, and complements the dynamical picture.

The RN solution provides a clean analytical example in which both signatures are present. In contrast, we show that regular black holes may exhibit neither, despite also having two horizons. In those cases the collapse is smooth and monotonic, and the surface approaches the center only asymptotically in proper time. We shall thus study the local and global criteria for these new features and their physical implications. Our work is also motivated by a desire to better understand black hole interiors, especially about the fate of the inner horizon, associated with Penrose's strong cosmic censorship conjecture (SCCC) \cite{sccc1,sccc2,Penrose:2002}. By identifying simple, geometry-based signatures in the OS framework, we offer a practical way to classify collapse outcomes into three distinct types: \emph{singular} collapse that ends at the center in finite proper time, \emph{bouncing} collapse that reverses at a finite radius, and \emph{soft-landing} collapse that approaches the center only asymptotically. In particular, the bouncing phenomenon indicates the instability of the inner horizon, consistent with the SCCC. We also emphasize that the collapse type can  be understood from the joint behavior of the evolutions of the star's surface $R(T)$ and the apparent horizon $R_{\rm AH}(T)$, which are physically linked via the proper time $T$ in the OS setup.

The paper is organized as follows. Section \ref{sec:setup} reviews the generalized OS collapse dynamics and presents the evolution equations for the star surface and horizons. Section \ref{sec:signatures} introduces the bounce and apparent-horizon minimum effects, explains how they lead to a simple classification of exterior spacetimes, and provides the conditions on the exterior metric that give rise to these features, using the Misner-Sharp mass function
\cite{MisnerSharp:1964}. Section \ref{sec:RN} applies this framework to the RN case, where both signatures are present. Section \ref{sec:regular} extends the analysis to regular black holes, offering a core-based classification of when apparent-horizon minima or bounces can occur and demonstrating that de~Sitter-core solutions exhibit neither feature.  We conclude in Section \ref{sec:conclusion}. We include some further material in Appendices. Appendix \ref{app:EMS} examines electrically-charged regular black holes in Einstein-Maxwell-scalar (EMS) gravity \cite{Li:2024rbw}. We show why neither signature appears on the de Sitter-core branch. Appendix \ref{sec:BI} further discusses the critical Born-Infeld exterior as an additional example with neither feature.

\section{OS collapse dynamics}
\label{sec:setup}

In the generalized OS framework \cite{OurPRD}, a collapsing spherical star is modeled as a homogeneous ball of pressureless dust, described by a FLRW spacetime in its interior. We consider the OS collapse with spatially-flat FLRW interior $(k=0)$, for which the interior geometry admits a convenient Painlev\'e-Gullstrand (PG)-type slicing. The star is matched at its surface to a special class of static, spherically symmetric exterior geometries of the form
\begin{equation}
ds^2=-f(r)\,dt^2+\frac{dr^2}{f(r)}+r^2 d\Omega^2.
\label{eq:static_metric}
\end{equation}
By ``special class'', we mean that $h=f$ in the general static metric \eqref{genstatic}. To avoid coordinate singularities on horizons and to work in a time slicing adapted to freely falling observers, we employ PG coordinates, where the interior FLRW metric can be written as
\begin{equation}
ds^2=-d\tau^2+\bigl(dr-rH(\tau)\,d\tau\bigr)^2+r^2 d\Omega^2.
\label{eq:FLRW_PG}
\end{equation}
In the above equation, $\tau$ is the proper time of comoving dust particles and $H(\tau)=\dot a/a$ is the Hubble parameter (dot denotes $d/d\tau$). On the star surface, the physical radius is $R(\tau)=a(\tau)\,r_0$, where $r_0$ is a fixed
comoving coordinate. In PG coordinates, the exterior becomes
\begin{equation}
ds^2=-dT^2+\bigl(dr+\sqrt{1-f(r)}\,dT\bigr)^2+r^2 d\Omega^2,
\label{eq:PG_metric}
\end{equation}
where $T$ is the PG time. Along the star surface, the interior proper time $\tau$ and the exterior PG time $T$ are identified, so we use a single time parameter $T$ throughout.

Imposing the Israel junction conditions (continuity of the induced metric and extrinsic curvature, with no thin shell) for a dust interior matched to \eqref{eq:PG_metric} yields the surface evolution equation
\begin{equation}
\dot R=\pm\sqrt{1-f(R)}\,,\qquad
H\equiv\frac{\dot R}{R}=\pm\frac{\sqrt{1-f(R)}}{R},
\label{eq:surface_kinematics}
\end{equation}
where an overdot denotes $d/dT$ and we take the collapsing star surface ($\dot R<0$). Before proceeding, we clarify two points regarding the scope and notation of our model. First, the symbol $R(\tau)$ (or $R(T)$) \textit{always} denotes the instantaneous radius of the collapsing star surface. It is never used as a general radial coordinate in the formulas for $\rho(\tau)$ and $p(\tau)$. Although the exterior is described by a charged spacetime, such as RN, we assume that the interior FLRW near the star surface consists only of \textit{neutral} dust in geodesic freefall. Its electromagnetic field extends into the exterior spacetime. The $R^{-4}$ scaling in the effective energy density arises from evaluating the electromagnetic field energy density $\rho_{\rm EM}=q^2/(8\pi r^4)$ at the moving boundary $r=R(\tau)$; this is a boundary condition imposed by the junction conditions, not a claim about the full interior field distribution \cite{OurPRD}. Second, the interior FLRW geometry is a kinematic ansatz, exactly as in the original OS model. We do not claim to construct a complete microphysical solution of the Einstein--Maxwell system with anisotropic interior matter. Instead, we prescribe a physically motivated exterior metric and use the Israel junction conditions to determine the unique interior evolution that matches it at the surface. The effective density $\rho(\tau)$ and pressure $p(\tau)$ are derived quantities, not assumed inputs. These expressions follow directly from substituting the surface evolution $\dot R=-\sqrt{1-f(R)}$ into the Friedmann equation for the interior FLRW metric. As shown explicitly in \cite{OurPRD}, these expressions satisfy the standard energy conditions for physically allowed ranges of $R(\tau)$; we verify this explicitly for the RN black hole. Furthermore, as we will show later, the star surface always bounces back at a finite radius without ever reaching the central region where the surface charge sources the exterior field. Consequently, the late-time collapse dynamics are completely determined by the exterior geometry $f(r)$ and are independent of the detailed interior charge distribution.

Integrating \eqref{eq:surface_kinematics} gives the star's surface evolution as
\begin{equation}
T(R)=-\int_{R_0}^{R}\frac{d\tilde R}{\sqrt{1-f(\tilde R)}}.
\label{eq:T_of_R}
\end{equation}
For the Schwarzschild exterior, corresponding to $f(r)=1-2m/r$, the above integration can be easily performed, and an analytic expression can be obtained. Reversing the function $T(R)$, we obtain the star-surface radius $R(T)$ as a function of its proper time $T$. However, for general $f(r)$, an analytic function $R(T)$ may not exist, and \eqref{eq:T_of_R} implies that it is more convenient to use the collapsing star-surface radius $R$ as a parameter to describe the dynamical collapse evolution. Using $R$ as if it were “time” is certainly allowed since $T(R)$ is a monotonically decreasing
function during the collapsing stage with shrinking $R$. In the OS collapse of the Schwarzschild black hole, $R$ ultimately shrinks to zero. However, as we shall demonstrate later for the RN case, the star surface will actually bounce back after $R$ reaches a certain minimum. After the bounce, we shall take the positive branch of eq.~(\ref{eq:surface_kinematics}) and hence the overall minus sign in (\ref{eq:T_of_R}) will reverse to plus sign. Consequently,  $T(R)$ becomes a monotonically increasing function.
	
During collapse, the boundary of the trapped region inside the star is the apparent horizon. Inside the FLRW interior, radial null geodesics satisfy $dr/dT = \pm 1 + rH$. The apparent horizon is located where outgoing null rays become trapped, which corresponds to the condition $1 + rH = 0$. Using the Hubble parameter $H = \dot R/R = -\sqrt{1-f(R)}/R$ from the surface evolution equation \eqref{eq:surface_kinematics}, we find that the interior apparent-horizon radius depends directly on the instantaneous star radius $R$:
\begin{equation}
	R_{\rm AH}(R)=\frac{R}{\sqrt{1-f(R)}}.
	\label{eq:RAH_def}
\end{equation}
This relation follows directly from the Friedmann equation and the Israel junction conditions, and it holds throughout the interior FLRW region. It establishes a precise link between the exterior metric function $f(r)$ and the interior trapped surface. Because of this link, any feature of the trapped surface---such as a bounce ($\dot R=0$, which requires $f(R_\ast)=1$) or a turning point in the apparent horizon ($dR_{\rm AH}/dT=0$, which requires $R f'(R) + 2(1-f(R))=0$)---is fixed by the behavior of $1-f(R)$ at the matching boundary. The interior trapped-surface evolution is therefore completely determined by the exterior geometry.

Let $R_+$ denote the outer horizon, defined as the largest positive root of $f(r)=0$. Since $f(R_+)=0$, Eq.~\eqref{eq:RAH_def} gives $R_{\rm AH}(R_+)=R_+$, meaning the apparent horizon coincides with the star surface at the moment it crosses the outer horizon. The same relation holds at the inner horizon $R_-$, where $R_{\rm AH}(R_-)=R_-$. The event horizon is obtained separately by tracing outgoing null rays backward in time, as described in Ref.~\cite{OurPRD}.

Together, Eqs.~\eqref{eq:T_of_R} and \eqref{eq:RAH_def} characterize both the proper time $T$ and apparent-horizon radius $R_{\rm AH}$ as functions of the star-surface radius $R$. These functions are completely determined by the exterior metric function $f(r)$. In the following sections, we use these expressions to reveal additional physically interesting signatures, namely bounce and apparent-horizon minimum, that can arise in the more general non-Schwarzschild exteriors that have two horizons.

\section{Bounce and apparent-horizon minimum}
\label{sec:signatures}

In this section we derive local, geometry-based criteria for the two possible signatures that could arise in generalized OS collapse: (i) a minimum of the apparent-horizon trajectory and (ii) a star-surface bounce. Both criteria can be determined directly from the exterior metric function $f(r)$.

\subsection{Apparent-horizon minimum}
\label{subsec:diag_vertex}

As the star collapses, the evolution of both the surface radius $R$ and the apparent-horizon radius $R_{\rm AH}$ can be expressed in terms of the time $T$. However, as mentioned under \eqref{eq:T_of_R}, $T$ is not the most convenient parameter for describing the evolution. Along the star's surface evolution, the apparent horizon is also naturally represented as the parametric curve $\bigl(R_{\rm AH}(R),\,T(R)\bigr)$, where $R$ denotes the instantaneous star-surface radius. The time derivative of the apparent horizon becomes
\begin{equation}
\frac{dR_{\rm AH}}{dT}=\frac{dR_{\rm AH}}{dR}\,\dot R\,.
\end{equation}
Using Eq.~\eqref{eq:surface_kinematics} on the collapsing branch ($\dot R<0$) and differentiating Eq.~\eqref{eq:RAH_def}, we find
\begin{equation}
\frac{dR_{\rm AH}}{dT}
= -\frac{G(R)}{2\bigl(1-f(R)\bigr)},
\qquad
G(R)\equiv R f'(R)+2\bigl(1-f(R)\bigr).
\label{eq:dRAH_dT}
\end{equation}
Along the physically allowed surface evolution one has $1-f(R)\ge0$, so the sign of
$dR_{\rm AH}/dT$ is governed by $G(R)$. The trajectory $R_{\rm AH}(T)$ has an extremum precisely, provided that
\begin{equation}
	1-f(R_{\rm turn})>0,\qquad G(R_{\rm turn})=0,\qquad
	G'(R_{\rm turn})>0,
	\label{eq:vertex_min_condition}
\end{equation}
where $R_{\rm turn}$ denotes the star-surface radius at which the apparent-horizon curve turns. The condition $G'(R_{\rm turn}) > 0$ ensures this is a local minimum. If $G'(R_{\rm turn}) < 0$, it would correspond to a local maximum, which is not the feature we study here. Since Eq.~\eqref{eq:dRAH_dT} and Eq.~\eqref{eq:vertex_min_condition} hold in a neighborhood of $R_{\rm turn}$, it follows that $dR_{\rm AH}/dT$ switches from negative to positive as $T$ increases. Therefore $R_{\rm AH}(T)$ has a \emph{local minimum}, which we call an \emph{apparent-horizon minimum}. Such non-monotonic behavior is absent in the Schwarzschild collapse but may arise in more general exteriors with two horizons, as we will demonstrate explicitly for the RN collapse in Section \ref{sec:RN}.

Using the Misner-Sharp mass function $m(R)$ \cite{MisnerSharp:1964},
\begin{equation}
f(R)=1-\frac{2m(R)}{R},
\label{eq:MS_mass_def}
\end{equation}
a turning point of $R_{\rm AH}(T)$ occurs at the star-surface radius $R=R_{\rm turn}$ when
\begin{equation}
\frac{dR_{\rm AH}}{dT}=0
\quad\Longleftrightarrow\quad
m'(R_{\rm turn})\,R_{\rm turn}-3m(R_{\rm turn})=0,
\qquad m(R_{\rm turn})>0.
\label{eq:vertex_condition}
\end{equation}
It is clear that such an apparent-horizon minimum is absent for the Schwarzschild exterior. We shall discuss the condition when it could arise and the constraints on the consistency of the generalized OS collapse.

\subsection{Star-surface bounce}
\label{subsec:diag_bounce}

On the collapsing OS surface ($\dot R<0$), Eq.~\eqref{eq:surface_kinematics} implies that the motion is real only when $f(R)\le 1$. It follows from \eqref{eq:surface_kinematics} that a turning point could occur when the collapse stops at some finite and nonzero $R_\ast$, namely
\begin{equation}
\dot R(R_\ast)=0
\qquad \Longleftrightarrow \qquad
f(R_\ast)=1.
\label{eq:turning_point}
\end{equation}
It is clear that such an $R_\ast$ does not exist in the Schwarzschild case. A \emph{ bounce} means that $R(T)$ attains a local minimum at $R_\ast$ and reverses into expansion, i.e. $\ddot R(R_\ast)>0.$
Differentiating~\eqref{eq:surface_kinematics} yields the exact identity
\begin{equation}
\ddot R
= -\frac{d}{dT}\sqrt{1-f(R)}
= -\frac{1}{2}\,f'(R),
\label{eq:ddot_general}
\end{equation}
so a turning point is a bounce precisely when
\begin{equation}
f(R_\ast)=1,
\qquad
f'(R_\ast)<0.
\label{eq:local_bounce}
\end{equation}
It is also useful to express the bounce criterion in terms of the Misner-Sharp mass function. The allowed region  becomes $m(R)\ge 0$, while the turning point \eqref{eq:turning_point} and the bounce condition \eqref{eq:local_bounce} imply the existence of $R_\ast$ such that
\begin{equation}
m(R_\ast)=0\,,\qquad m'(R_\ast)>0\,.
\label{eq:mass_turn}
\end{equation}
Therefore, a surface bounce can occur only if the exterior develops an inner forbidden pocket with $f(R)>1$ (or $m(R)<0$) for $0<R<R_\ast$, which we refer to as an RN-type repulsive core. As we shall see later, the requirement that $m'(R_\ast)>0$ is ensured by the null energy condition (NEC).

There is also another special radius during the collapse that is worth noting. Before the surface stops and reverses at $R_\ast$, there must already exist an \emph{inflection radius} $R_{\rm infl}$ where the collapse changes from accelerating
to decelerating. This inflection point satisfies
\begin{equation}
\ddot R(R_{\rm infl})=0 \quad\Longleftrightarrow\quad f'(R_{\rm infl})=0,
\end{equation}
with $\dot R(R_{\rm infl})\ne 0$. At $R_{\rm infl}$ the surface is
still collapsing, but its acceleration switches sign. Note that this inflection radius $R_{\rm infl}$ is generally different from the apparent-horizon turning point $R_{\rm turn}$ defined by \eqref{eq:vertex_condition}.

Considering the case where the exterior spacetime admits exactly two horizons, \(0 < R_- < R_+\), such that \(f(R_\pm) = 0\), we suppose the OS surface undergoes a bounce at \(R_\ast > 0\), meaning that \(R_\ast\) is the inner boundary of the allowed region and satisfies
	\begin{equation}
		f(R_\ast) = 1, \qquad f(R) > 1 \quad \text{for all} \quad 0 < R < R_\ast .
	\end{equation}
	Since we restrict to spacetimes with two horizons in this discussion,  \(f(R)\) does not cross zero again for \(R < R_-\). For a bounce to occur, the region \(R < R_\ast\) must be classically forbidden, meaning \(f(R) > 1\) for all \(0 < R < R_\ast\). If we assumed \(R_\ast \ge R_-\), then \(R_-\) would lie in this forbidden region, implying \(f(R_-) > 1\). This contradicts the definition of the inner horizon, \(f(R_-) = 0\). Therefore, the bouncing point must lie strictly inside the inner horizon, i.e.,
	\begin{equation}
		R_\ast < R_-.
\end{equation}
It is worth commenting on the number of horizons a black hole can have. It was shown in literature \cite{HorizonLimit:2021} that SEC prohibits more than two horizons. Multiple horizons can arise if we relax the SEC. Black holes with four horizons satisfying DEC were constructed \cite{Liu:2019rib}. However, for simplicity, we shall only consider spacetimes with no more than two horizons in this paper.

\subsection{Interplay between the bounce and the apparent-horizon minimum}
\label{subsec:interplay_bounce_vertex}

The local criteria for an apparent-horizon minimum \eqref{eq:vertex_condition} and a surface bounce \eqref{eq:local_bounce} are independent. However, in a standard asymptotically flat OS collapse starting from a sufficiently large initial radius, a bounce enforces the existence of an apparent-horizon minimum along the collapsing star surface, but the converse may not be true.

\paragraph{Bounce $\Rightarrow$ an apparent-horizon minimum.} Along the collapsing surface, the apparent-horizon slope is given by,
\begin{equation}
\frac{dR_{\rm AH}}{dT}
=\frac{m'(R)\,R-3m(R)}{2m(R)}\,.
\label{eq:interplay_slope}
\end{equation}
Asymptotic flatness implies $m(R)\to M$ and $m'(R)\to0$ as $R\to\infty$, hence
$\frac{dR_{\rm AH}}{dT}\to -\frac{3}{2}<0$ at early stages of collapse.
If the surface undergoes a bounce at $R_\ast>0$, namely $m(R_\ast)=0$ with $m'(R_\ast)>0$, then as $R\to R_\ast^+$ the numerator approaches $m'(R_\ast)R_\ast>0$ while the denominator $2m(R)\to0^+$, so $\frac{dR_{\rm AH}}{dT}\to +\infty$. By continuity, there must exist a $R_{\rm turn}\in(R_\ast,R_0)$ such that
\begin{equation}
\frac{dR_{\rm AH}}{dT}\Big|_{R_{\rm turn}}=0,
\end{equation}
which yields a minimum on the collapsing star surface.

\paragraph{Apparent-horizon minimum without bounce.} Conversely, an apparent-horizon minimum may occur without any bounce.
It suffices to choose an asymptotically flat mass profile with $m(R)>0$ for all $R>0$, while $m'(R)R-3m(R)$ changes sign. A simple analytic example is
\begin{equation}
m(R)=M\Bigl(1-e^{-(R/L)^p}\Bigr),
\qquad M>0,\quad L>0,\quad p>3.
\end{equation}
It is clear that  $m(R)>0$ for all $R>0$ and $m(R)\to M$ as $R\to\infty$, so no bounce occurs. Meanwhile, the quantity $m'(R)R-3m(R)=M\bigl[(p x+3)e^{-x}-3\bigr]$ with $x=(R/L)^p$ is positive for small $R$ (since $p>3$) and it approaches $-3M$ as $R\to\infty$. Thus, the quantity vanishes at some finite $R_{\rm turn}>0$.
By \eqref{eq:interplay_slope}, this produces an apparent-horizon minimum of $R_{\rm AH}(T)$.

\section{RN collapse: a bounce with an apparent-horizon minimum}
\label{sec:RN}

The RN geometry provides a simple explicit example of a static exterior that can exhibit \emph{both} signature features discussed in Section~\ref{sec:signatures}: a bounce of the star surface and an apparent-horizon minimum in the apparent-horizon
trajectory. The RN exterior spacetime is described by
\begin{equation}
f_{\rm RN}(r)=1-\frac{2m}{r}+\frac{q^{2}}{r^{2}},
\label{eq:RN_f}
\end{equation}
where $m$ and $q$ denote the mass and the electric charge. The corresponding Misner--Sharp mass function is
\begin{equation}
m(r)=m-\frac{q^{2}}{2r}.
\label{eq:RN_mR}
\end{equation}
For $0<|q|<m$ there are two horizons,
\begin{equation}
R_{\pm}=m\pm\sqrt{m^{2}-q^{2}},
\label{eq:RN_horizons}
\end{equation}
with $R_{+}$ the radius of the outer horizon and $R_{-}$ the inner horizon; the extremal limit $|q|=m$ gives $R_{+}=R_{-}=m$. Throughout this section we restrict to the black hole parameter $0<|q|\le m$ and assume the OS collapse starts from a sufficiently large initial radius $R_0>R_+$.

\subsection{Apparent-horizon minimum}
\label{subsec:RN_vertex}

\begin{figure}[t]
  \centering
  \begin{subfigure}{0.49\linewidth}
    \centering
    \includegraphics[width=\linewidth]{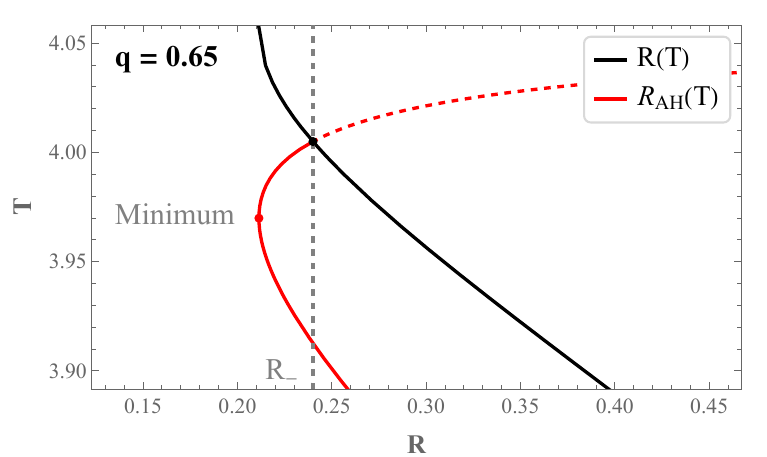}
    \caption{Disallowed ordering: $T(R_{\rm turn})<T(R_-)$.\label{fig:RN_bad_order}}
  \end{subfigure}
  \hfill
  \begin{subfigure}{0.49\linewidth}
    \centering
    \includegraphics[width=\linewidth]{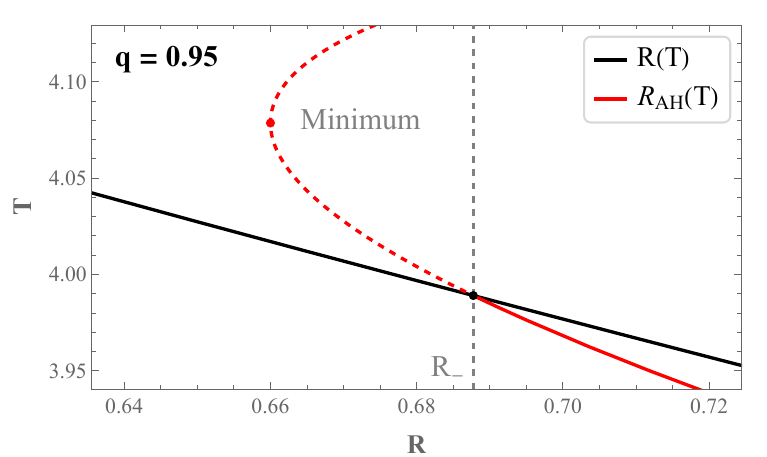}
    \caption{Allowed ordering: $T(R_{\rm turn})\ge T(R_-)$.\label{fig:RN_good_order}}
  \end{subfigure}
  \caption{\small Depending on the charges, two possible time orderings between the apparent-horizon turning point and the inner-horizon crossing
  in RN OS collapse with $m=1$ (left: $q=0.65$; right: $q=0.95$).
  The black curve is the star-surface trajectory $(R,\,T(R))$, while the red curve is the apparent-horizon trajectory
  $\bigl(R_{\rm AH}(R),\,T(R)\bigr)$.
  The condition \eqref{eq:RN_vertex_physcrit} is a global consistency requirement:
  it ensures that the relevant inner apparent-horizon evolution connects smoothly to the surface crossing of the inner
  horizon; otherwise, the turning occurs too early, while the surface is still outside $R_-$.}
  \label{fig:RN_vertex_time_order}
\end{figure}

Substituting \eqref{eq:RN_mR} into the general apparent-horizon minimum condition \eqref{eq:vertex_condition} yields
\begin{equation}
R_{\rm turn}=\frac{2q^{2}}{3m}.
\label{eq:RN_RV}
\end{equation}
This apparent-horizon minimum is dynamically accessible since
\begin{equation}
1-f_{\rm RN}(R_{\rm turn})=\frac{3m^{2}}{4q^{2}}>0.
\end{equation}
For a physically consistent formation of a two-horizon black hole in the GOS framework, the inner apparent horizon must connect the outer and inner horizons as the star's surface crosses them. In the $(R,T)$ diagram, the apparent-horizon trajectory is the parametric curve $\bigl(R_{\rm AH}(R),\,T(R)\bigr)$, so the apparent-horizon minimum is located at $\bigl(R_{\rm AH}(R_{\rm turn}),\,T(R_{\rm turn})\bigr)$,
where $R_{\rm turn}$ is determined by $dR_{\rm AH}/dT=0$. The inner-horizon crossing of the surface occurs at the point $\bigl(R_-,\,T(R_-)\bigr)$. During collapse, the apparent horizon marks the boundary of the trapped region inside the star. By definition, all particles and null rays inside this region move inward. For the matching to remain physically consistent, this boundary must shrink monotonically as the surface falls. If $R_{\rm turn} > R_-$, the apparent horizon would reach a minimum and begin to expand while the star surface is still outside the inner horizon. This contradicts the monotonic shrinking required for a trapped-surface boundary in a collapsing configuration. The nontrivial requirement is therefore the correct \emph{time ordering}: the apparent-horizon minimum must occur no earlier than the crossing of the apparent horizon over the inner horizon,
\begin{equation}
	T(R_{\rm turn})\ge T(R_-).
	\label{eq:RN_vertex_timecrit}
\end{equation}
On the collapsing star surface, we have $\frac{dT}{dR}=-\frac{1}{\sqrt{1-f(R)}}<0$, so $T(R)$ increases monotonically as $R$ decreases. Hence \eqref{eq:RN_vertex_timecrit} is equivalent to the simple radial condition
\begin{equation}
	R_{\rm turn}\le R_-,
	\label{eq:RN_vertex_physcrit}
\end{equation}
which ensures that the apparent horizon evolution remains consistent with the definition of a trapped region until the surface crosses $R_-$. This is a geometric consistency condition of the GOS framework, not an arbitrary restriction imposed by the exterior. Here $R_\pm=m\pm\sqrt{m^{2}-q^{2}}$ are the RN horizon radii. Together with $q^{2}\le m^{2}$, we obtain the consistent range on the charge $q$
\begin{equation}
\frac{\sqrt{3}}{2}\,m\le |q|\le m.
\label{eq:RN_charge_interval}
\end{equation}
In particular, when $q^{2}=3m^{2}/4$, we have $R_{\rm turn}=R_-$. In Fig.~\ref{fig:RN_vertex_time_order}, we illustrate the consistent and inconsistent apparent-horizon minimum structures in the OS collapse. The left panel corresponds to a small charge where $R_{\rm turn} > R_-$. In this case, the inner apparent horizon begins to expand, which violates the definition of the apparent horizon as the boundary of a trapped surface, where no outgoing null geodesics should exist. Therefore, consistency requires $R_{\rm turn} \le R_-$, which gives $|q| \ge \sqrt{3}m/2$.

The requirement $|q| \ge \sqrt{3}m/2$ introduces a discontinuity as $q \to 0$. This does not mean weakly charged black holes cannot form in nature. It indicates that the perfectly symmetric OS setup becomes mathematically inconsistent for small charge. This discontinuity is a geometric signal of the well-known instability of the Cauchy horizon. In a realistic collapse, even infinitesimal perturbations would trigger mass inflation or destroy the inner horizon long before the symmetric OS trajectory reaches the inconsistent regime. The breakdown for small $q$ is therefore consistent with Penrose's strong cosmic censorship conjecture, which states that Cauchy horizons are unstable and cannot persist in generic collapse. The idealized OS model ``detects'' this instability through the breakdown of trapped-region consistency.

\subsection{Star-surface bounce}
\label{subsec:RN_bounce}

In the RN black hole, the charge contribution produces an inner repulsive barrier, so the OS surface can reach a \emph{bounce radius}. On the collapsing star surface, the bounce radius $R_\ast$ is determined by $f_{\rm RN}(R_\ast)=1$, which gives
	\begin{equation}
		R_\ast=\frac{q^{2}}{2m}.
		\label{eq:Rstar_RN}
	\end{equation}
The bounce radius in Eq.~\eqref{eq:Rstar_RN} is determined solely by the exterior RN metric function $f_{\rm RN}(r)$. Since the surface bounces at $R_\ast > 0$, the OS trajectory never probes the central singularity at $R=0$. Consequently, within the present matching framework, the bounce is determined by the exterior geometry evaluated at the stellar surface, without requiring any specification of the charge distribution inside the FLRW bulk. The FLRW interior remains neutral, and the charge parameter $q$ is carried entirely by the boundary surface $\Sigma$. To confirm that $R_\ast$ is a bounce, we use the general relation in Eq.~\eqref{eq:ddot_general}. For the RN metric,
\begin{equation}
f_{\rm RN}'(R_\ast)
=-\frac{8m^{3}}{q^{4}}<0
\quad\Longrightarrow\quad
\ddot R(R_\ast)=\frac{4m^{3}}{q^{4}}>0,
\end{equation}
so the surface reaches $R_\ast$ with vanishing velocity and positive outward acceleration, and therefore rebounds.
This bounce occurs for all $0<|q|\le m$, and one has the ordering
\begin{equation}
R_\ast<R_{-}\le R_{+}\,.
\end{equation}
Thus we see that in the OS framework, the star surface never reaches $R=0$; instead, it bounces back at $R_\ast < R_-$ and accumulates near $R_-$. This behavior is consistent with SCCC \cite{Penrose:2002}, which conjectures that the inner horizon of the RN is unstable and should not remain regular in any matter perturbation, possibly via the phenomenon of mass inflation, where even tiny perturbations grow exponentially near the inner horizon, potentially creating a curvature singularity there \cite{Poisson:1990eh}. Our result supports this picture: the bounce prevents the surface from reaching the central singularity, effectively ending the classical evolution near the inner horizon instead. 

\subsubsection{Surface trajectory and near-bounce behavior}
\label{subsec:RN_TR}

On the collapsing star surface, the OS surface satisfies Eq.~\eqref{eq:surface_kinematics}, hence
\begin{equation}
\frac{dT}{dR}=-\frac{1}{\sqrt{1-f_{\rm RN}(R)}}
=-\frac{R}{\sqrt{2mR-q^{2}}}\,.
\label{eq:RN_dT_dR}
\end{equation}
Imposing the initial condition $T(R_0)=0$, the surface worldline in the $(R,T)$ plane can be written schematically as
\begin{equation}
T(R)=
\begin{cases}
\frac{(mR_0+q^2)\sqrt{2mR_0-q^2}-(mR+q^2)\sqrt{2mR-q^2}}{3m^2},
\qquad &(R_\ast\le R\le R_0),  \quad \text{collapse}\,\\[2pt]
2T_\ast - \frac{(mR_0+q^2)\sqrt{2mR_0-q^2}-(mR+q^2)\sqrt{2mR-q^2}}{3m^2}, & (R_\ast \le R\le R_-), \quad \text{bounce}\,
\end{cases}
\label{eq:RN_Tcol_Texp}
\end{equation}
where $T_\ast\equiv T(R_\ast)$ denotes the proper time at the bounce radius $R_\ast=q^2/(2m)$.
The two branches join smoothly at the bounce point $(R_\ast,T_\ast)$.

Moreover, the bounce is locally regular. Expanding $f_{\rm RN}(r)$ around $r=R_\ast$ gives
$1-f_{\rm RN}(R)\simeq -f'_{\rm RN}(R_\ast)(R-R_\ast)$ with $f'_{\rm RN}(R_\ast)<0$, so that the surface radius has a
quadratic minimum,
\begin{equation}
R(T)=R_\ast+\frac{2m^{3}}{q^{4}}\,(T-T_{\ast})^{2}
+\mathcal O\!\bigl((T-T_{\ast})^{4}\bigr).
\label{eq:RN_quadratic_minimum}
\end{equation}

After the surface crosses the outer horizon $R_+$, it enters the trapped region $R_-<R<R_+$, where the causal
	structure prevents any outward escape back to the original exterior. In the idealized OS evolution on a fixed RN
	background one may formally continue the trajectory to $R<R_-$ and reach the bounce radius $R_\ast$, but in a realistic
	collapse with full nonlinear effects taken into account,  this continuation is not expected to be physically meaningful: the star effectively gets ``captured'' by the
	inner horizon. This is consistent with the well-known instability of the inner horizon due to mass inflation~\cite{Poisson:1990eh}, as discussed above. Nevertheless, it is of interest to see that the simple OS model appears also to reveal the possible instability of the inner horizon of the RN black hole. It should also be noted that the consistency interval \eqref{eq:RN_charge_interval} ensures that the bounce radius $R_\ast$ is not parametrically small and is therefore a
	physically meaningful feature.

\begin{figure}[t]
\centering
\includegraphics[width=0.47\textwidth]{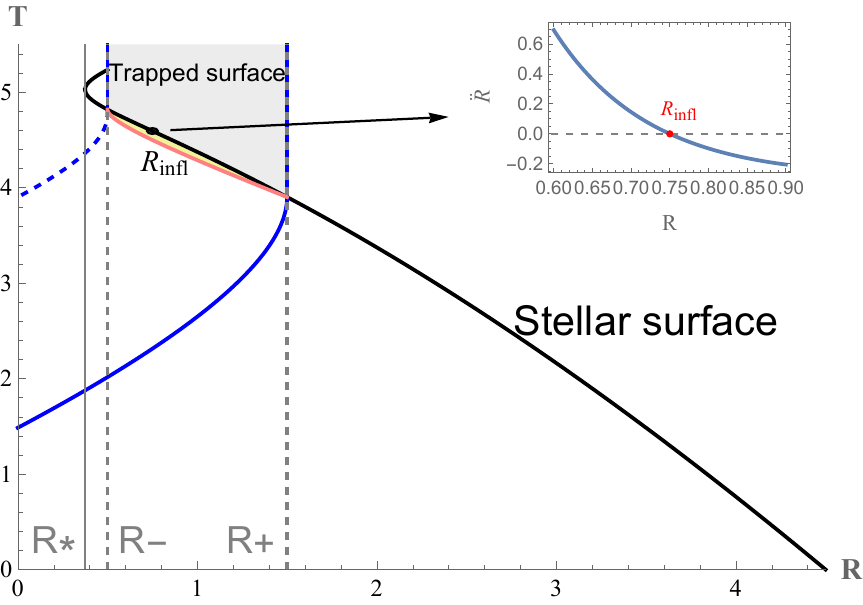}
\caption{\small RN collapse in the $(R,T)$ plane.
	The initial radius is $R_0=4.5$, and the inner and outer horizons are set to $(R_{-},R_{+})=(0.5,\,1.5)$.
	These choices fix the mass and charge as $m=(R_{+}+R_{-})/2=1$ and $q^2=R_{+}R_{-}=0.75$.
	The collapse features three key radii: the surface bounce radius $R_\ast=0.375$, the apparent-horizon minimum $R_{\rm turn}=0.5=R_{-}$, and the inflection point $R_{\rm infl}=0.75$. The black curve shows the star surface $R(T)$.
	The red and blue curves denote the apparent horizon and event horizon, respectively. The trapped regions are shaded.}
\label{fig:RN_TR}
\end{figure}

To complete the picture of RN collapse, we also consider the \emph{inflection point} $R_{\rm infl}$, where the surface
acceleration vanishes, i.e.\ $\ddot R=0$, which gives
\begin{equation}
R_{\rm infl}=\frac{q^2}{m}.
\end{equation}
Note that $R_{\rm infl}=2R_\ast$, so it lies outside the bounce radius. At this point, the surface is still collapsing,
with $\dot R(R_{\rm infl})=-m/|q|\ne0$. However, the acceleration changes sign: collapse \emph{speeds up} for $R>R_{\rm infl}$
($\ddot R<0$) and \emph{slows down} for $R<R_{\rm infl}$ ($\ddot R>0$) until the bounce occurs at $R_\ast$.
The apparent-horizon minimum radius $R_{\rm turn}=2q^2/(3m)$ lies between the bounce and inflection points,
\begin{equation}
R_\ast=\frac{q^2}{2m} \;<\; R_{\rm turn}=\frac{2q^2}{3m} \;<\; R_{\rm infl}=\frac{q^2}{m}.
\end{equation}
In Fig.~\ref{fig:RN_TR}, we draw the diagram of OS collapse of the RN black holes with sufficiently large charge so that the apparent-horizon minimum does not occur before the surface crosses the inner horizon. We illustrate the inflection radius, after which the star surface decelerates and then bounces back to the inner horizon. The matter accumulation in the inner horizon indicates that it is unstable, consistent with Penrose's SCCC. 

\FloatBarrier
\section{Regular black hole collapse}
\label{sec:regular}

In the previous sections, we have seen that the OS collapse of black holes with an additional inner horizon such as the RN black hole can have new features, such as the bounce of the star surface inside the inner horizon and also the possible existence of the apparent-horizon minimum of the apparent horizon. However, having two horizons is only a necessary, but not sufficient condition for these new features. Regular black holes necessarily have two horizons, but the analysis of the OS collapse of the Bardeen black hole illustrates that there is neither bounce nor apparent-horizon minimum \cite{ShojaiSadeghiHassannejad:2022}. In this section, we show that this is in fact generally true for regular black holes. It turns out that this observation allows us to give a new proof that regular black hole with a Minkowski core necessarily violates the NEC.

\subsection{Regular black holes: no bounce or apparent-horizon minimum}

A regular black hole with the special static metric \eqref{eq:static_metric} is characterised by a regular core, which can be de Sitter (dS), anti-de Sitter (AdS) or Minkowski. For these spacetimes, the same logic applies: the surface evolution is governed by the exterior metric function $f(r)$ evaluated at the star surface, and the interior regular core only matters insofar as it ensures $f(r)$ has the required behavior near $R=0$. Since the surface approaches $R=0$ only asymptotically in the soft-landing case, the detailed core structure does not affect the qualitative collapse outcome. It was shown that such regular black holes satisfying the NEC must have a de Sitter core \cite{Li:2023yyw}. Requiring NEC also implies that the regular black hole must have $f(r)\le 1$, with inequality saturated at both asymptotic infinity $(r\rightarrow \infty)$ and at the core ($r=0$). It is then clear that there can be no bounce, since a necessary condition for bounce is the existence of a finite $R_\ast$ such that $f(R_\ast)=1$. Furthermore, as the collapsing star-surface radius $R$ approaches the core asymptotically, namely
\be
R\sim e^{-T/\tau_0}\,,\qquad \tau_0 = \sqrt{\ft3{\Lambda_{\rm eff}}}\,,\label{tau0}
\ee
where $\Lambda_{\rm eff}$ is the effective cosmological constant of the dS core. The asymptotic behavior yields what we call a {\it soft-landing} collapsing scenario. An explicit concrete example is given in Appendix \ref{app:EMS}.

In the previous RN example, we have also seen that an apparent-horizon minimum can arise.
However, when the charge is too small, the apparent-horizon minimum lies inside the star, making the OS collapse inconsistent. This inconsistency disappears for sufficiently large charge. We now show that apparent-horizon minimum will never arise in the OS collapse of regular black holes satisfying the NEC. To see this, we consider the exterior metric profile \eqref{eq:static_metric} and define the Misner-Sharp mass $m(r)$.
We require that the exterior matter satisfies the NEC, which for \eqref{eq:static_metric} implies
\begin{equation}
\frac12 f''(r)+\frac{1-f(r)}{r^2}\ge0.
\label{eq:NEC_ineq_f}
\end{equation}
In Misner-Sharp language, this becomes
\begin{equation}
\frac12 f''(r)+\frac{1-f(r)}{r^2}
=\frac{2m'(r)-r\,m''(r)}{r^2}\ge0
\qquad\Longleftrightarrow\qquad
\frac{d}{dr}\!\left(\frac{m'(r)}{r^{2}}\right)\le0,
\label{eq:NEC_monotone_mprime}
\end{equation}
so $m'(r)/r^2$ is nonincreasing outward. Suppose in addition that the center is regular in the sense that
\begin{equation}
m(0)=0, \qquad m(r)=\mathcal O(r^3)\qquad (r\to0).
\label{eq:regular_center_assumption}
\end{equation}
Then for any $r>0$ and all $0<x<r$ one has
\begin{equation}
\frac{m'(x)}{x^2}\ge \frac{m'(r)}{r^2}
\qquad\Longrightarrow\qquad
m'(x)\ge \frac{m'(r)}{r^2}\,x^2.
\end{equation}
Integrating from $x=0$ to $x=r$ yields the inequality
\begin{equation}
m(r)=\int_0^r m'(x)\,dx
\ge \frac{m'(r)}{r^2}\int_0^r x^2\,dx
=\frac{r\,m'(r)}{3},
\qquad\Longrightarrow\qquad
3m(r)-r\,m'(r)\ge0.
\label{eq:3m_rmprime_nonneg}
\end{equation}
Recalling $G(r)=r f'(r)+2(1-f(r))=\frac{2}{r}\bigl(3m(r)-r\,m'(r)\bigr)$, we obtain $G(r)\ge0$.
Along the physically allowed collapsing star surface ($1-f(R)\ge0$), Eq.~\eqref{eq:dRAH_dT} gives
\begin{equation}
\frac{dR_{\rm AH}}{dT}
=-\frac{G(R)}{2(1-f(R))}\le0.
\label{eq:no_vertex_ds_nec}
\end{equation}
Therefore, $R_{\rm AH}(T)$ is monotone and cannot develop an apparent-horizon minimum. Note that this conclusion relies crucially on regularity at the center and does not apply to singular exteriors such as RN, for which $m(r)$ is not integrable at $r=0$ and $3m-rm'$ can change sign.

\subsection{Regular core-based classification}
\label{subsec:regular_core_based}

We now classify regular-core exteriors by their near-origin expansion of $f(r)$ and discuss for each core type whether an apparent-horizon minimum or a bounce is expected. For the special static metric of the type \eqref{eq:static_metric}, there can be three different types of regular cores: dS, AdS and Minkowski cores. The AdS core requires SEC, in which case, it can be shown that the core cannot be extended to asymptotic Minkowski infinity \cite{Li:2023yyw}. We thus consider the dS core:
\begin{equation}
f(r)=1-\frac{\Lambda_{\rm eff}}{3}\,r^2+\mathcal O(r^4)\qquad (r\to0),\qquad (\Lambda_{\rm eff}>0).
\end{equation}
We have
\begin{equation}
m(r)=\frac{r}{2}\bigl(1-f(r)\bigr)=\frac{\Lambda_{\rm eff}}{6}\,r^3+\mathcal O(r^5),
\qquad
m'(r)=\frac{\Lambda_{\rm eff}}{2}\,r^2+\mathcal O(r^4).
\end{equation}
Hence $m'(r)r-3m(r)=\mathcal O(r^5)$ and Eq.~\eqref{eq:interplay_slope} gives $dR_{\rm AH}/dT\to 0$ as $R\to0$.
This near-core flattening alone does not exclude a finite-radius turning point.
However, if the exterior satisfies the NEC on the relevant accessible region, then
we have shown that $dR_{\rm AH}/dT\le 0$ throughout the collapse and thus excludes any minimum. In fact, it is easy to verify that $R_{\rm AH}$ approach to the constant $\tau_0$, given in \eqref{tau0}, asymptotically in large proper time $T$. We illustrate these features in the OS collapse of the electrically-charged regular black hole of EMS gravity in Appendix \ref{app:EMS}.

Regular black holes with a Minkowski core, on the other hand, \emph{have at least one minimum} provided the collapsing surface can probe sufficiently small radii while remaining in the allowed region ($m>0$). This does not generically produce a bounce. For example, considering
\begin{equation}
f(r)=1-c_4 r^4+\mathcal O(r^6)\qquad (r\to0),\qquad (c_4>0),
\end{equation}
we have $m(r)=\tfrac{c_4}{2}r^5+\mathcal O(r^7)$ and $m'(r)=\tfrac{5c_4}{2}r^4+\mathcal O(r^6)$, hence
\begin{equation}
m'(r)\,r-3m(r)=c_4 r^5+\mathcal O(r^7)>0
\quad\Longrightarrow\quad
\frac{dR_{\rm AH}}{dT}\to 1>0\qquad (R\to0).
\end{equation}
On the other hand, asymptotic flatness implies $m(R)\to M$ and $m'(R)\to0$ as $R\to\infty$, so
$\frac{dR_{\rm AH}}{dT}\to -\tfrac32<0$ at early stages. Therefore $\frac{dR_{\rm AH}}{dT}$ must vanish at least once at some intermediate $R_{\rm turn}$ by continuity, yielding an apparent-horizon minimum of $R_{\rm AH}(T)$ on the infalling evolution. Although we only considered a specific example to illustrate the point, the statement is generally true for all regular black holes with a Minkowski core. 

In Section 5.1, we showed that any regular black hole satisfying NEC cannot have an apparent-horizon minimum. On the other hand, we showed here that a regular spacetime with Minkowski core always produces an apparent-horizon minimum. Combining these two  statements, we conclude that regular black holes with Minkowski core necessarily violate the NEC. This gives a new proof of the theorem in \cite{Li:2023yyw}, using collapse dynamics instead of direct energy condition analysis.

\section{Conclusion}
\label{sec:conclusion}

In this paper, we studied the generalized OS collapse of non-Schwarzschild black holes, focusing on those with two horizons, inner Cauchy and outer event horizons. The OS collapse with the FLRW interior requires the exterior metric to be a special static metric of the type \eqref{eq:static_metric}. We found that new features could arise when a black hole has two horizons. The first is \emph{an apparent-horizon minimum} in the apparent-horizon evolution. Along the collapsing surface, the apparent-horizon radius $R_{\rm AH}(T)$ can have a temporary minimum. This behavior is controlled by the local turning-point conditions in Eqs.~\eqref{eq:dRAH_dT}-\eqref{eq:vertex_min_condition} (or equivalently Eq.~\eqref{eq:vertex_condition}).  The second is a \emph{bounce} at a nonzero radius of the star's surface. In terms of $f(r)$, this occurs when $f(R_\ast)=1$ with $f'(R_\ast)<0$ (Eq.~\eqref{eq:local_bounce}), ensuring that the turning point is a local minimum of $R(T)$. Equivalently, the Misner-Sharp mass $m(r)$ crosses zero from below, indicating an inner repulsive barrier that reflects the collapsing surface. Such a bounce is excluded in Schwarzschild metric and, more generally, in exteriors with $m(r)\ge 0$ for
all $r>0$. We also highlighted an effect, the \emph{inflection radius} $R_{\rm infl}$ defined by $\ddot R(R_{\rm infl})=0$, where the collapse switches from accelerating to decelerating while remaining on the collapsing star surface, before the bounce takes place.

For the RN exterior, we explicitly showed that the OS collapse exhibits both new features.
Requiring the apparent-horizon turning point to lie inside the inner horizon ($R_{\rm turn}\le R_-$) forces the charge to
be sufficiently large, which also ensures that the bounce scale is not parametrically small. Thus, RN collapse provides a
clean analytic benchmark featuring three distinct radii:
\begin{itemize}
\item $R_\ast=q^2/(2m)$: the surface stops and bounces ($\dot R=0$, $\ddot R>0$);
\item $R_{\rm infl}=q^2/m$: the collapse changes from accelerating to decelerating ($\dot R\ne0$, $\ddot R=0$);
\item $R_{\rm turn}=2q^2/(3m)$: the apparent horizon reaches a minimum (see condition \eqref{eq:vertex_condition}).
\end{itemize}

However, the existence of both inner and outer horizons does not necessarily imply that the two new features must occur in the OS collapse. In fact, for regular black holes, which necessarily have two horizons, their OS collapse has no apparent-horizon minimum, nor any bounce. Instead, the collapse exhibits a soft landing: the surface approaches $R\to 0$ only asymptotically in proper time, while $R_{\rm AH}(R)$ tends to a finite limiting radius. Furthermore, the NEC-based monotonicity result for regular centers implies a monotone apparent-horizon trajectory and excludes any minimum. This allows us to prove a no-go theorem that regular black holes with Minkowski core necessarily violate the NEC. We explicitly illustrate the OS collapse of regular black holes in EMS in Appendix \ref{app:EMS}. As another explicit example, we showed in Appendix \ref{sec:BI} that the critical Einstein-Born-Infeld exterior also exhibits neither a bounce nor an apparent-horizon minimum.

We now summarize the qualitative behaviors of the star surface $R(T)$ in the OS collapse for black holes with at most two horizons. The collapse leads to one of three possible outcomes:
\begin{itemize}
	\item \textbf{Singular collapse (e.g.\ Schwarzschild):}
	The surface reaches $R=0$ in finite proper time. After trapped surfaces form, the apparent-horizon radius
	$R_{\rm AH}(T)$ decreases monotonically and terminates at the center at the final collapse time.
	\item \textbf{Bouncing collapse (e.g.~RN):}
	The surface stops at a finite radius $R_\ast$ and reverses into expansion.
	In a standard asymptotically flat OS evolution starting from sufficiently large $R_0$,  a bounce may be accompanied by a turning behavior of $R_{\rm AH}(T)$; in fact, under standard asymptotically flat OS initial data, a bounce implies at least one apparent-horizon minimum (see Sec.~\ref{subsec:interplay_bounce_vertex}).
	Both the surface and the apparent horizon show this turning behavior because of the same repulsive effect in the spacetime.
	\item \textbf{Soft-landing collapse (e.g.~regular black holes):}
	The star never reaches the center in finite time; it only gets closer and closer as time goes to infinity. The apparent horizon also changes smoothly and never turns around. Neither the surface nor the horizon shows a bounce or an apparent-horizon minimum.
\end{itemize}
These three classes are summarized in Table~\ref{tab:classification}.
\begin{table}[h]
	\centering
	\caption{Summary of OS collapse outcomes for different exterior spacetimes.}
	\label{tab:classification}
	\renewcommand{\arraystretch}{1.3}
	\begin{tabular}{lccc}
		\hline
		\textbf{Exterior Model} & \textbf{Correction Term} & \textbf{Bounce?} & \textbf{Collapse Type} \\
		\hline
		Schwarzschild & None & No & Singular \\
		Reissner-Nordstr\"om & $+q^2/r^2$ & Yes & Bouncing \\
		Regular BH (dS core) & Phenomenological/UV & No & Soft-landing \\
		qOS (LQC) & $+\alpha m^2/r^4$ & Yes & Bouncing \\
		\hline
	\end{tabular}
\end{table}
Thus, by inspecting how the star surface and the apparent horizon behave, we can tell which type of collapse is actually occurring. It is useful to compare this classification with the quantum OS construction of Ref.~\cite{Lewandowski:2023}. In that work, the effective LQC dynamics of the dust interior leads, through the junction conditions, to a quantum-deformed Schwarzschild exterior with a $1/r^4$ correction. Applying our criteria to the qOS metric, $f_{\text{qOS}}(r) = 1 - 2m/r + \alpha m^2/r^4$, yields a nonzero bounce radius $R_* = (\alpha m/2)^{1/3}$, an apparent-horizon turning scale $R_{\text{turn}} = (\alpha m)^{1/3}$, and an inflection scale $R_{\text{infl}} = (2\alpha m)^{1/3}$. Thus, the quantum OS model belongs to the bouncing class in our geometric classification. However, the physical origin of the bounce differs from that in the RN case. In the RN collapse, the repulsive effect is generated by the exterior electric charge, whereas in the quantum OS model, it arises from the LQC corrections to the effective dust dynamics.

Our results may provide an insight into the (in)stability of the inner horizon that is relevant to SCCC. For the RN black hole with a timelike singularity, the bounce of the OS collapse indicates that matter accumulates on the horizon and hence it is unstable. For regular black holes, the existence of the inner horizon may not violate SCCC, and interestingly, there will be no bounce in the OS collapse.

In this paper, we considered special static black holes with at most two horizons. We found that three outcomes could arise in OS collapse. Black holes with multiple horizons beyond two could arise when the SEC is violated \cite{Liu:2019rib}. It is of great interest to investigate the fate of the OS collapse in these more complicated cases.

\section*{Acknowledgements}

We are grateful to Fatimah Shojai for useful discussions. This work is supported in part by the National Natural Science Foundation of China (NSFC)
grants No. W2533015, No. 12375052 and No. 11935009, as well as by the Tianjin University
Self-Innovation Fund Extreme Basic Research Project Grant No. 2025XJ21-0007.

\begin{appendices}

\section{OS collapse of regular EMS black holes}
\label{app:EMS}

Since Bardeen constructed the first example of the regular black hole metric, there have been efforts in constructing fundamental theories that admit such regular black holes. Notable examples include Einstein gravity coupled to the nonlinear electrodynamics (e.g.~\cite{ABG:1998,ABG:1999,Fan:2016hvf,Li:2023yyw}) and quasi-topological pure gravity theory, e.g.~\cite{Bueno:2024dgm}. In this appendix, we present an explicit example of the OS collapse of electrically-charged regular black holes in EMS gravity. As in the RN example, the charge parameter $q$ characterizes the exterior electromagnetic field. In the corresponding GOS collapse framework, the FLRW bulk interior is assumed to be neutral and homogeneous. Consequently, the exterior charge is supported entirely by an idealized surface charge on $\Sigma$. We also present a representative collapse diagram, illustrating explicitly the absence of both a surface bounce and an apparent-horizon minimum, in agreement with the discussion in Section \ref{sec:regular}.

We consider the EMS theory with action
\begin{equation}
S=\frac{1}{16\pi}\int d^4x\,\sqrt{-g}\left( R-\phi^{-1}\mathcal F - V(\phi) \right),
\label{eq:EMS_action}
\end{equation}
where $\mathcal F\equiv \tfrac14 F_{\mu\nu}F^{\mu\nu}$ and $F_{\mu\nu}=\partial_\mu A_\nu-\partial_\nu A_\mu$.
The scalar $\phi$ is auxiliary (no kinetic term); its algebraic equation of motion relates $\phi$ to $\mathcal F$. Eliminating $\phi$ yields an effective nonlinear electrodynamics in the usual $F$-framework. The auxiliary scalar field in \eqref{eq:EMS_action} is convenient for organizing solution branches and imposing regularity at the core.

\begin{figure}[t]
\centering
\includegraphics[width=0.45\textwidth]{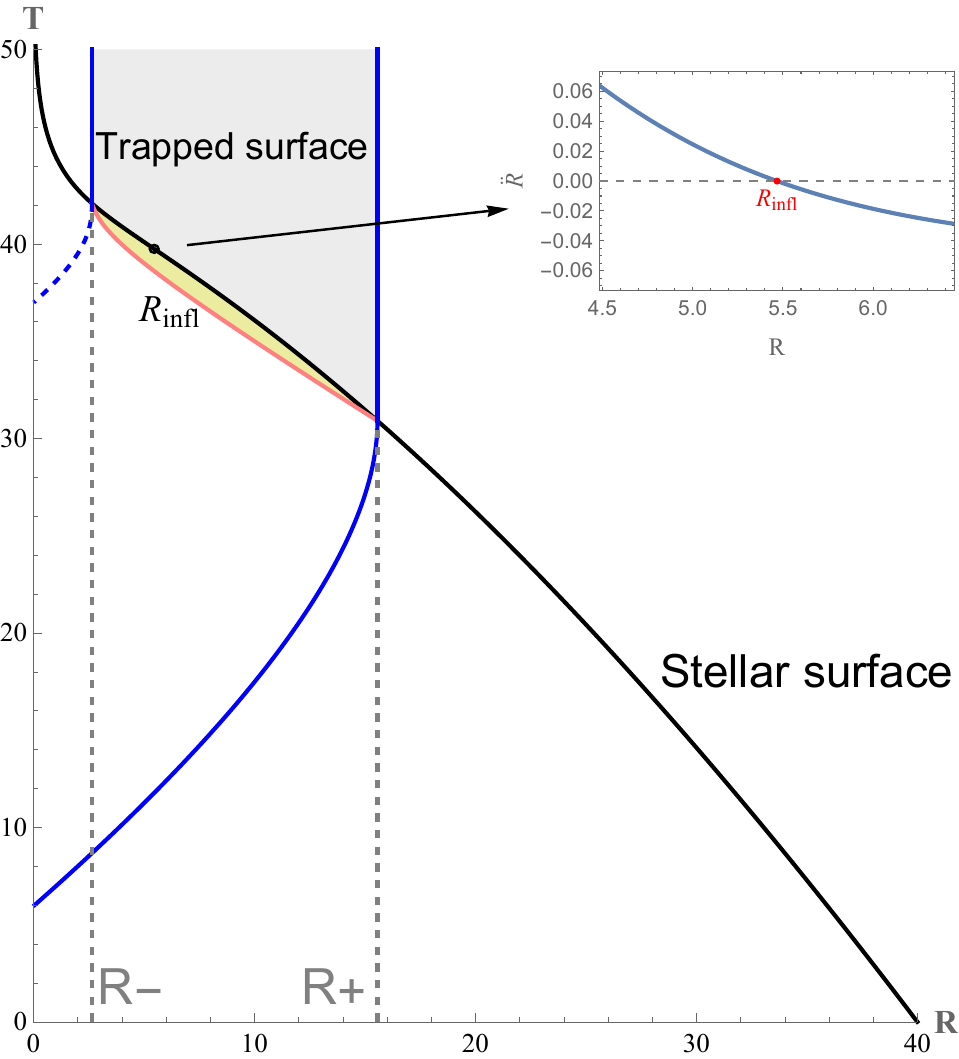}
\caption{\small Collapse diagram in the $(R,T)$ plane for an electrically-charged regular black hole in EMS gravity with $(n,\alpha)=(1,1)$ and $(q,m)=(15,m_{\rm cr}(15))$, starting from $R_0=40$. The vertical lines indicate the horizon radii $R_\pm$ in \eqref{eq:EMS_Rpm_numeric}. The star surface $R(T)$ (black) exhibits no turning point and thus no bounce. The apparent-horizon curve $R_{\rm AH}(T)$ (pink) is monotone and has no apparent-horizon minimum. The inflection radius $R_{\rm infl}\simeq 5.47$ is defined by $\ddot R=0$: the collapse accelerates in the $R>R_{\rm infl}$ region and decelerates in the $R<R_{\rm infl}$ region, consistent with the soft-landing behavior as $R\to0$.}
\label{fig:EMS_TR}
\end{figure}

As an explicit solvable example we take the $V_1$ family~\cite{Li:2024rbw}. For general $(n,\alpha)$ one has
\begin{align}
V_1(\phi) &= \frac{1}{\alpha}\Bigl(1-\phi^{\frac{n}{n+1}}\Bigr)^{\frac{n+1}{n}}, \label{eq:EMS_V1_app}\\
f(r) &= 1-\frac{2m}{r}+\frac{q^2}{4r^2}\,
{}_2F_1\!\left(\frac{1}{4n},\frac{1}{n};1+\frac{1}{4n};
-\left(\frac{\alpha q^2}{2r^4}\right)^n\right).
\label{eq:EMS_f_V1_app}
\end{align}
The solution involves two independent parameters, the mass $m$ and the charge $q$. The solution is generally singular, but regularity at the origin can be realized on the critical branch, namely $m=m_{\rm cr}(q)$, where the would-be $1/r$ term cancels in the
small-$r$ expansion and the geometry develops a dS core. For instance, for $(n,\alpha)=(1,1)$ and $q=15$ we find numerically $m=m_{\rm cr}(15)\approx 9.592$, and the resulting geometry has two horizons at
\begin{equation}
R_-\simeq 2.655,
\qquad
R_+\simeq 15.574.
\label{eq:EMS_Rpm_numeric}
\end{equation}
In this example one verifies numerically that $f(r)<1$ for all $r>0$, so no surface turning point satisfying $f(R_\ast)=1$ with $f'(R_\ast)<0$ exists (hence no bounce). Moreover, the effective density $\rho_{\rm ext}(r)$ decreases outward on $0<r<R_+$, which implies $dR_{\rm AH}/dT<0$ along the collapsing branch and therefore we have a monotone apparent-horizon curve with no apparent-horizon minimum, consistent with the discussion in Section \ref{sec:regular}. Relevant diagram is plotted in Fig.~\ref{fig:EMS_TR}. In particular, we see explicitly that there exists an inflection point where the collapse transitions from the accelerating phase to the decelerating phase. Nevertheless, there is no bounce of the star surface, which monotonically collapses to the core only asymptotically in the proper time.

\section{OS collapse of critical Born-Infeld black hole}
\label{sec:BI}

We show that the Einstein-Born-Infeld (EBI) exterior, when in critical mass/charge relation, provides another concrete example of a non-Schwarzschild spacetime with \emph{neither} an apparent-horizon minimum nor a bounce. For a purely electric EBI black hole the theory is of $L(\mathcal F)$ type \cite{BornInfeld:1934,Dey:2004},
\begin{equation}
L(\mathcal{F})
=\frac{4}{\alpha}\left(1-\sqrt{1+\frac{\alpha}{2}\,\mathcal{F}}\right),
\qquad \alpha>0,
\end{equation}
where $\alpha$ sets the Born-Infeld scale. The theory admits charged black hole solution of the ansatz \eqref{eq:static_metric}, with $f$ given by
\be
f_{\rm BI}=1 - \frac{2 M}{r} + \frac{2 r^2}{3 \alpha}\left(1 - {}_2F_1\!\left[-\frac{3}{4}, -\frac{1}{2}, \frac{1}{4}, -\frac{\alpha q^2}{4 r^4}\right]\right).
\ee
For generic mass and charge $(M,q)$, the small-$r$ expansion of $f_{\rm BI}(r)$ contains a $1/r$ term. The metric has time-like singularity, as in the case of the RN black hole. Thus in the general cases, there will be both bounce and apparent-horizon minimum.

We are interested in the case where the $1/r$ term vanishes by requiring its cancellation selects a distinguished critical mass $M=M_{\rm cr}$, where
\begin{equation}
M_{\rm cr}
=\frac{q^{3/2}\,\Gamma\!\left(\frac14\right)^2}
{12\sqrt{2\pi}\,\alpha^{1/4}}\,,
\label{eq:BI_Mcr}
\end{equation}
for which the core is less singular. On this critical branch the near-origin expansion reads
\begin{equation}
f_{\rm BI}(r)
=1-\frac{q}{\sqrt{\alpha}}+\frac{2}{3\alpha}\,r^{2}+\mathcal O(r^{4}),
\qquad (r\to0),
\label{eq:BI_core_expansion}
\end{equation}
so in general $f_{\rm BI}(0)=1-q/\sqrt{\alpha}\neq 1$.
In Misner--Sharp language one finds
\begin{equation}
m(r)=\frac{r}{2}\bigl(1-f_{\rm BI}(r)\bigr)
=\frac{q}{2\sqrt{\alpha}}\,r-\frac{1}{3\alpha}\,r^{3}+\mathcal O(r^{5}),
\qquad (r\to0),
\label{eq:BI_m_smallr}
\end{equation}
hence $m(r)>0$ for sufficiently small $r>0$, and the OS reality condition $m(R)\ge0$ holds near the center.

\paragraph{No apparent-horizon minimum and no bounce.}
By the apparent-horizon minimum criterion derived in Section~\ref{sec:signatures}, a minimum can occur only if the apparent-horizon slope along the collapsing surface vanishes at some accessible radius. On the critical EBI branch this never happens. Indeed, the regular-core expansion \eqref{eq:BI_core_expansion} implies $G(r)=\frac{2q}{\sqrt{\alpha}}+\mathcal O(r^4)>0$ as $r\to0^+$, while asymptotic flatness gives $G(r)=\frac{6M_{\rm cr}}{r}-\frac{q^2}{r^2}+\cdots>0$ as $r\to\infty$. Using the explicit critical EBI profile, one furthermore verifies that $G(r)$ remains strictly positive for all $r>0$ on the accessible branch. Since $1-f(r)>0$ holds along the collapse, Eq.~\eqref{eq:dRAH_dT} then yields $dR_{\rm AH}/dT<0$ throughout the evolution, so $R_{\rm AH}(T)$ is monotone and admits no apparent-horizon minimum.

A bounce requires an inner forbidden pocket $f(r)>1$ and a first turning point $R_\ast>0$ where $f(R_\ast)=1$ with $f'(R_\ast)<0$; see Section~\ref{subsec:diag_bounce}. This is impossible on the critical EBI branch because $f_{\rm BI}(r)\le 1$ holds throughout the accessible region. Therefore the collapsing surface encounters no turning point and no bounce occurs.

\end{appendices}

\end{document}